\documentclass[onecolumn,amsmath,showpacs,nofootinbib,12pt]{revtex4-2}
\usepackage{graphicx}
\usepackage{dcolumn}
\usepackage{bm}
\usepackage{color} 
\usepackage{slashed}
\begin{document}
\newcommand{\hs}{\hspace*{0.5cm}}
\newcommand{\vs}{\vspace*{0.5cm}}
\newcommand{\be}{\begin{equation}}
\newcommand{\ee}{\end{equation}}
\newcommand{\bea}{\begin{eqnarray}}
\newcommand{\eea}{\end{eqnarray}}
\newcommand{\ben}{\begin{enumerate}}
\newcommand{\een}{\end{enumerate}}
\newcommand{\bde}{\begin{widetext}}
\newcommand{\ede}{\end{widetext}}
\newcommand{\nn}{\nonumber}
\newcommand{\crn}{\nonumber \\}
\newcommand{\Tr}{\mathrm{Tr}}
\newcommand{\non}{\nonumber}
\newcommand{\noi}{\noindent}
\newcommand{\al}{\alpha}
\newcommand{\la}{\lambda}
\newcommand{\bet}{\beta}
\newcommand{\ga}{\gamma}
\newcommand{\va}{\varphi}
\newcommand{\om}{\omega}
\newcommand{\pa}{\partial}
\newcommand{\+}{\dagger}
\newcommand{\fr}{\frac}
\newcommand{\bc}{\begin{center}}
\newcommand{\ec}{\end{center}}
\newcommand{\Ga}{\Gamma}
\newcommand{\de}{\delta}
\newcommand{\De}{\Delta}
\newcommand{\ep}{\epsilon}
\newcommand{\varep}{\varepsilon}
\newcommand{\ka}{\kappa}
\newcommand{\La}{\Lambda}
\newcommand{\si}{\sigma}
\newcommand{\Si}{\Sigma}
\newcommand{\ta}{\tau}
\newcommand{\up}{\upsilon}
\newcommand{\Up}{\Upsilon}
\newcommand{\ze}{\zeta}
\newcommand{\ps}{\psi}
\newcommand{\Ps}{\Psi}
\newcommand{\ph}{\phi}
\newcommand{\vph}{\varphi}
\newcommand{\Ph}{\Phi}
\newcommand{\Om}{\Omega}
\newcommand{\AdrHEPC}{Phenikaa Institute for Advanced Study and Faculty of Basic Science, Phenikaa University, Yen Nghia, Ha Dong, Hanoi 100000, Vietnam}

\title{Interpreting dark matter solution for $B-L$ gauge symmetry} 

\author{Phung Van Dong} 
\email{dong.phungvan@phenikaa-uni.edu.vn}
\affiliation{\AdrHEPC} 

\date{\today}

\begin{abstract}

It is shown that the solution for $B-L$ gauge symmetry with $B-L=-4,-4,+5$ assigned for three right-handed neutrinos respectively, reveals a novel scotogenic mechanism with implied matter parity for neutrino mass generation and dark matter stability. Additionally, the world with two-component dark matter is hinted.   
                  
\end{abstract} 

\maketitle

{\it Introduction}.---Of the exact conservations in physics, the conservation of baryon number minus lepton number, say $B-L$, causes curiosity. There is no necessary principle for $B-L$ conservation since it results directly from the standard model gauge symmetry. As a matter of fact, every standard model interaction separately preserves $B$ and $L$ such that $B-L$ is conserved and anomaly-free, if three right-handed neutrinos, say $\nu_{1R},\nu_{2R},\nu_{3R}$, are simply imposed. In the literature, there are two integer solutions for $B-L$, such as $-1,-1,-1$ and $-4,-4,+5$, according to $\nu_{1R},\nu_{2R},\nu_{3R}$ \cite{blnumber}. In contrast to electric and color charges, the excess of baryons over antibaryons of the universe suggests that $B-L$ is broken. $B-L$ likely occurs in left-right symmetry and grand unification, which support the first solution and a seesaw mechanism for neutrino mass generation \cite{Minkowski,Yanagida,Gell-Mann,Mohapatra}, but no such traditional theories manifestly explain the existence of dark matter, similarly to the standard model. 

With regard to the second solution, Refs. \cite{addbl1,addbl2,Okada} discussed type-I seesaw neutrino mass generation, in which the first work interpreted $\nu_{3R}$ dark matter by including a $Z_2$ symmetry, the second work studied a scalar dark matter by choosing an alternative $Z_2$, while the third work investigated $\nu_{3R}$ dark matter stability without needing any extra symmetry as $Z_2$. The accidental stability of a scalar dark matter was also probed in Ref. \cite{Singirala} along with neutrino mass generation by an effective interaction. Alternatively, Ref. \cite{addbl5} discussed Dirac or inverse seesaw neutrino mass generation by imposing extra vectorlike leptons $N$'s with $B-L=-1$ and interpreting $\nu_{1,2R}$ dark matter, and subsequently Ref. \cite{addbl6} gave a realization of residual $Z_3$ symmetry and a long-lived scalar dark matter in such a combined framework. Ref. \cite{addblno} discovered a simple option of Dirac neutrino mass as suppressed by a potential with accidental $\nu_{3R}$ dark matter. Ref. \cite{addbl7} investigated radiative neutrino mass and $\nu_{3R}$ dark matter mass by introducing an extra vectorlike lepton $S$ with $B-L=8$ besides $\nu_R$'s, whereas Ref. \cite{Ma} proposed a scotogenic scheme in a variant with seven extra neutral leptons alternative to $\nu_R$'s. Ref. \cite{addbl9} examined radiative Dirac neutrino mass and dark matter by imposing extra lepton doublets and $Z_2$ for dark matter stability.     

As a next attempt to the above process, I argue that the second solution provides naturally both dark matter and neutrino mass, without requiring any extra fermion and extra symmetry. It is indeed the first scotogenic mechanism realized for minimal right-handed neutrino content with $B-L=-4,-4,+5$ and a residual matter parity, alternative to~\cite{scoto}. In a period, the matter parity which stabilizes dark matter has been found usefully in supersymmetry. I argue that the matter parity naturally arises from the second solution for $B-L$ gauge symmetry, without necessity of supersymmetry.  

\begin{table}[h]
\bc
\begin{tabular}{lccccc}
\hline\hline 
Field & $SU(3)_C$ & $SU(2)_L$ & $U(1)_Y$ & $U(1)_{B-L}$ & $P$\\
\hline
$l_{aL} = \begin{pmatrix}
\nu_{aL}\\
e_{aL}\end{pmatrix}$ & 1 & 2 & $-1/2$ & $-1$ & $+$\\
$\nu_{\al R}$ & 1 & 1 & 0 & $-4$ & $-$\\
$\nu_{3R}$ & 1 & 1 & 0 & $5$ & $+$\\
$e_{aR}$ & 1 & 1 & $-1$ & $-1$ & $+$\\
$q_{aL}= \begin{pmatrix}
u_{aL}\\
d_{aL}\end{pmatrix}$ & 3 & 2 & $1/6$ & $1/3$ & $+$\\
$u_{aR}$ & 3 & 1 & $2/3$ & $1/3$ & $+$\\
$d_{aR}$ & 3 & 1 & $-1/3$ & $1/3$ & $+$\\
$H=\begin{pmatrix}
H^+\\
H^0\end{pmatrix}$ & 1 & 2 & $1/2$ & 0 & $+$\\  
$\phi_1$ & 1 & 1 & 0 & 8 & $+$\\
$\phi_2$ & 1 & 1 & 0 & $-10$ & $+$\\
$\chi =\begin{pmatrix}
\chi^0\\
\chi^-\end{pmatrix}$ & 1 & 2 & $-1/2$ & 3 & $-$\\  
$\eta$ & 1 & 1 & 0 & 5 & $-$\\
\hline\hline 
\end{tabular}
\caption[]{\label{tab1} Field presentation content of the model.}
\ec
\end{table}    

\begin{figure}[h]
\bc
\includegraphics[scale=1]{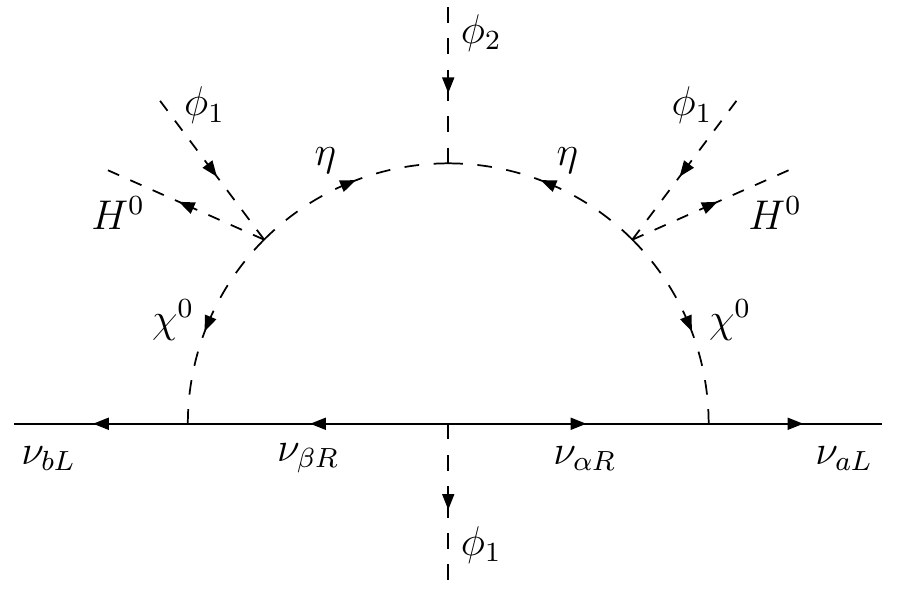}
\caption[]{\label{fig1} Neutrino mass generation induced by dark matter solution of $B-L$ gauge symmetry.}
\ec
\end{figure}

{\it Proposal}.---Gauge symmetry is given by $SU(3)_C\otimes SU(2)_L\otimes U(1)_Y\otimes U(1)_{B-L}$. Field content according to this symmetry is supplied in Tab.~\ref{tab1}, in which $a=1,2,3$ and $\al=1,2$ indicate family indices. The usual Higgs field $H$ has a VEV $\langle H \rangle =(0,v/\sqrt{2})$ breaking the electroweak symmetry and generating mass for usual particles. The new Higgs fields $\phi_{1,2}$ have VEVs, $\langle \phi_1\rangle = w_1/\sqrt{2}$ and $\langle \phi_2\rangle = w_2/\sqrt{2}$, inducing Majorana masses for $\nu_{\al R}$ and $\nu_{3R}$ respectively as well as breaking $B-L$, determining a residual matter parity $P=(-1)^{3(B-L)+2s}$ (see below), which is included to Tab. \ref{tab1} too. I impose $w_{1,2}\gg v=246$ GeV for consistency with the standard model. Additionally, the scalar $\chi$ couples $l_{aL}$ to $\nu_{\al R}$, while the scalar $\eta$ couples $\chi$ to $H\phi_1$ as well as to $\phi_2$, which radiatively generates neutrino mass (see Fig. \ref{fig1}). The fields $\chi,\eta$ have vanished VEVs, preserved by the matter parity conservation. This realizes a scotogenic scheme with automatic matter parity by the model itself, which stabilizes dark matter candidates $\nu_{\al R}, \chi^0,\eta$, opposite to \cite{scoto} for which a $Z_2$ is {\it ad hoc} input.    

{\it Matter parity}.---A $B-L$ transformation has the form, $P=e^{ix(B-L)}$, where $x$ is a parameter. $P$ conserves both the vacua $w_{1,2}$, i.e. $Pw_1=w_1$ and $Pw_2=w_2$, given that $e^{i8x}=1$ and $e^{-i 10 x}=1$. It leads to $x=k\pi$, thus $P=(-1)^{k(B-L)}$, for $k$ integer. Acting $P$ on every field, I derive $P=1$ for minimal $|k|=6$, except the identity with $k=0$. This defines a residual group $Z_6=\{1,p,p^2,p^3,p^4,p^5\}$, where $p=(-1)^{B-L}$ and $p^6=1$. I factorize $Z_6=Z_2\otimes Z_3$, where $Z_2=\{1,p^3\}$ is the invariant subgroup of $Z_6$, while $Z_3=\{[1],[p^2],[p^4]\}$ is the quotient group of $Z_6$ by $Z_2$, with each coset element containing two elements of $Z_6$, i.e. $[g]=\{g,gp^3\}$, thus $[1]=[p^3]=Z_2$, $[p^2]=[p^5]=\{p^2,p^5\}$, and $[p^4]=[p]=\{p,p^4\}$. Since $[p^4]=[p^2]^2=[p^2]^*$ and $[p^2]^3=[1]$, $Z_3$ is generated by the generator $[p^2]=[\om^{3(B-L)}]$, where $\om =e^{i 2\pi/3}$ is the cube root of unity. Since $3(B-L)$ is integer due to $p^6=1$, $Z_3$ has three irreducible representations $\underline{1}$, $\underline{1}'$, and $\underline{1}''$ according to $[p^2]=[1]\to 1$, $[p^2]=[\om]\to \om$, and $[p^2]=[\om^2]\to \om^2$ respectively, which are homomorphic from those of $Z_6$ independent of the signs $p^3=\pm 1$ that identify $Z_6$ elements in a coset \cite{matterparity}. I obtain $[p^2]=[\om]\rightarrow \om\sim \underline{1}'$ for quarks, while $[p^2]=[1]\rightarrow 1\sim \underline{1}$ for all other fields. Hence, $Z_3$ transforms nontrivially only for quarks, isomorphic to the center of the color group. In other words, the theory automatically conserves $Z_3$, accidentally preserved by $SU(3)_C$. Omitting $Z_3$, what the residual symmetry remains is only $Z_2=\{1,p^3\}$, generated by the generator $p^3=(-1)^{3(B-L)}$. Since the spin parity $p_s=(-1)^{2s}$ is always conserved by the Lorentz symmetry, I redefine \be P\equiv p^3\times p_s=(-1)^{3(B-L)+2s},\label{mpd}\ee to be matter parity similar to that in supersymmetry, governing this model.\footnote{The matter parity appeared in previous studies, e.g., \cite{Ma}.} The matter-parity group $M=\{1,P\}$ instead of $Z_2$ has two irreducible representations $\underline{1}$ and $\underline{1}'$ according to $P=1$ and $P=-1$ respectively, collected in Tab. \ref{tab1} for every field. The lightest of odd fields $\nu_{\al R}, \eta,\chi$ is absolutely stabilized by the matter parity conservation, providing a dark matter candidate. However, since $\nu_{3R}$ does not singly couple to standard model fields at renormalizable level similar to proton, $\nu_{3R}$ has a lifetime bigger than the universe age (see below), supplying an alternative dark matter candidate, kind of minimal dark matter.   

{\it Scalar potential and mass splitting}.---I write the scalar potential $V=V(H,\phi_1,\phi_2)+V(\eta,\chi,\mathrm{mix})$, where the first part includes only the fields that induce breaking,
\bea V(H,\phi_1,\phi_2) &=& \kappa^2 H^\dagger H + \kappa^2_1 \phi^*_1 \phi_1 + \kappa^2_2 \phi^*_2 \phi_2 + c (H^\dagger H)^2 + c_1 (\phi^*_1 \phi_1)^2 + c_2 (\phi^*_2 \phi_2)^2\crn
&& + c_3 (H^\dagger H) (\phi^*_1 \phi_1) + c_4 (H^\dagger H)(\phi^*_2 \phi_2)+c_5 (\phi^*_1 \phi_1) (\phi^*_2 \phi_2),\eea while the second part is relevant to $\eta,\chi$ and mixed terms with breaking fields, \bea V(\eta,\chi,\mathrm{mix})&=& \mu^2_1 \eta^* \eta + \mu^2_2 \chi^\dagger \chi + \la_1 (\eta^* \eta)^2 +\la_2 (\chi^\dagger \chi)^2 +\la_3 (\eta^*\eta)(\chi^\dagger\chi)\crn 
&&+(\eta^*\eta)(\la_{4} \phi^*_1\phi_1+\la_{5}\phi^*_2\phi_2 +\la_6 H^\dagger H)+(\chi^\dagger \chi)(\la_{7} \phi^*_1\phi_1+\la_{8}\phi^*_2\phi_2 +\la_9 H^\dagger H)\crn
&&+\la_{10}(\chi^\dagger H)(H^\dagger \chi)+\mu (\phi_2 \eta \eta +H.c.)+\la (H\chi \eta \phi^*_1+H.c.).\eea The trivial $\eta,\chi$ and nontrivial $H,\phi_{1,2}$ vacua acquire $\mu^2_{1,2}>0$, $\kappa^2<0$, and $\kappa^2_{1,2}<0$. Additionally, the potential bounded from below demands that $c>0$, $c_{1,2}>0$, and $\la_{1,2}>0$, which are derived from $V>0$ when $H$, $\phi_{1}$, $\phi_{2}$, $\eta$, and $\chi$ separately tending to infinity. Further, $V>0$ applies when every two of these fields simultaneously tend to infinity, yielding $c_3>-2\sqrt{cc_1}$, $c_4>-2\sqrt{cc_2}$, $c_5>-2\sqrt{c_1c_2}$, $\la_3>-2\sqrt{\la_1 \la_2}$, $\la_4>-2\sqrt{\la_1 c_1}$, $\la_5>-2\sqrt{\la_1 c_2}$, $\la _6>-2\sqrt{\la_1 c}$, $\la_7>-2\sqrt{\la_2 c_1}$, $\la_8 >-2\sqrt{\la_2 c_2}$, and $\la_{9}+\la_{10}\Theta(-\la_{10})>-2\sqrt{\la_2 c}$, where $\Theta$ is the Heaviside step function. Notice that $V>0$ for every three (every four, every five) of scalar fields simultaneously tending to infinity will supply extra, complicated conditions for scalar self-couplings.\footnote{All such conditions ensure the quartic coupling matrix to be co-positive responsible for the vacuum stability, which can be derived with the aid of \cite{copositivity}.} Furthermore, constraints of physical scalar masses squared to be positive might exist, but most of which would be equivalent to the given conditions.  

Let $\chi^0=(R+iI)/\sqrt{2}$ and $\eta=(R_1+i I_1)/\sqrt{2}$. Further, denote $M^2_1=\mu^2_1+\fr{\la_4}{2}w^2_1+\fr{\la_5}{2}w^2_2+\fr{\la_6}{2}v^2$ and $M^2_2=\mu^2_2+\fr{\la_7}{2}w^2_1+\fr{\la_8}{2}w^2_2+\fr{\la_9}{2}v^2$, which all are at least at $w_{1,2}$ scale. The field $\chi^\pm$ is a physical field by itself with mass $m^2_{\chi^\pm}=M^2_2+\fr{\la_{10}}{2}v^2$. The fields $R,R_1$ and $I,I_1$ mix in each pair, such as
\bea V &\supset& \fr 1 2 \begin{pmatrix} R_1 & R \end{pmatrix} \begin{pmatrix} M^2_1+\sqrt{2}\mu w_2 & -\fr 1 2 \la v w_1\\
-\fr 1 2 \la v w_1 & M^2_2 \end{pmatrix} \begin{pmatrix} R_1\\ R\end{pmatrix}\crn
&&+\fr 1 2 \begin{pmatrix} I_1 & I \end{pmatrix} \begin{pmatrix} M^2_1-\sqrt{2}\mu w_2 & \fr 1 2 \la v w_1\\
\fr 1 2 \la v w_1 & M^2_2 \end{pmatrix} \begin{pmatrix} I_1\\ I\end{pmatrix}. \eea I define two mixing angles,\be t_{2\theta_R}=\fr{- \la v w_1}{M^2_2-M^2_1-\sqrt{2}\mu w_2},\hs t_{2\theta_I}=\fr{ \la v w_1}{M^2_2-M^2_1+\sqrt{2}\mu w_2}.\ee The physical fields are 
\bea && R'_1=c_{\theta_R} R_1 - s_{\theta_R} R,\hs R'=s_{\theta_R} R_1 + c_{\theta_R} R, \\ 
&& I'_1=c_{\theta_I} I_1 - s_{\theta_I} I,\hs I'=s_{\theta_I} I_1 + c_{\theta_I} I, \eea with respective masses,
\bea && m^2_{R'_1}\simeq M^2_1+\sqrt{2}\mu w_2+\fr{\fr 1 4 \la^2 v^2 w^2_1}{M^2_1+\sqrt{2}\mu w_2-M^2_2},\hs m^2_{R'}\simeq M^2_2+\fr{\fr 1 4 \la^2 v^2 w^2_1}{M^2_2-M^2_1-\sqrt{2}\mu w_2},\\
&&m^2_{I'_1}\simeq M^2_1-\sqrt{2}\mu w_2+\fr{\fr 1 4 \la^2 v^2 w^2_1}{M^2_1-\sqrt{2}\mu w_2-M^2_2},\hs m^2_{I'}\simeq M^2_2+\fr{\fr 1 4 \la^2 v^2 w^2_1}{M^2_2-M^2_1+\sqrt{2}\mu w_2}, \eea where the approximations come from $|\theta_{R,I}|\ll 1$ due to $v\ll w_{1,2}$, and it is clear that the $R,I$ and $R_1,I_1$ masses are now separated. 

{\it Neutrino mass}.---The Yukawa Lagrangian relevant to neutral fermions is 
\be \mathcal{L}_{\mathrm{Yuk}}\supset h_{a\al}\bar{l}_{aL} \chi \nu_{\al R}+\fr 1 2 t_{\al \beta} \bar{\nu}^c_{\al R} \nu_{\beta R}\phi_1+\fr 1 2 t_{33} \bar{\nu}^c_{3R}\nu_{3R}\phi_2+H.c.\ee When $\phi_{1,2}$ develop VEVs, $\nu_{R}$'s obtain Majorana masses, such as \be m_{\nu_{1R}}=-t_{11} \fr{w_1}{\sqrt{2}},\hs m_{\nu_{2R}}=-t_{22}\fr{w_1}{\sqrt{2}},\hs m_{\nu_{3R}}=-t_{33}\fr{w_2}{\sqrt{2}},\ee where I assume $t_{\al \beta}$ to be flavor diagonal, i.e. $\nu_{1,2,3R}$ are physical fields by themselves. This Yukawa Lagrangian combined with the above scalar potential, i.e. $\mathcal{L} \supset \fr{h_{a\al }}{\sqrt{2}} \bar{\nu}_{aL}(c_{\theta_R} R'+ic_{\theta_I} I'-s_{\theta_R} R'_1-is_{\theta_I} I'_1)\nu_{\al R} -\fr 1 2 m_{\nu_{\al R}} \nu^2_{\al R}+H.c. -\fr 1 2 m^2_{R'} R'^2-\fr 1 2 m^2_{I'} I'^2-\fr 1 2 m^2_{R'_1}R'^2_1-\fr 1 2 m^2_{I'_1}I'^2_1$, up to kinetic terms yields necessary features for the diagram in Fig. \ref{fig1} in mass basis. That said, the loop is propagated by physical fermions $\nu_{1,2R}$ and physical scalars $R',I',R'_1,I'_1$, inducing neutrino mass in form of $\mathcal{L}\supset -\fr 1 2 \bar{\nu}_{aL} (m_{\nu})_{ab}\nu^c_{bL}+H.c.$, where
\bea (m_\nu)_{ab} &=& \fr{h_{a\al}h_{b\al}m_{\nu_{\al R}}}{32\pi^2}\left(\fr{c^2_{\theta_R} m^2_{R'}\ln \fr{m^2_{\nu_{\al R}}}{m^2_{R'}}}{m^2_{\nu_{\al R}}-m^2_{R'}}-\fr{c^2_{\theta_I}m^2_{I'}\ln \fr{m^2_{\nu_{\al R}}}{m^2_{I'}}}{m^2_{\nu_{\al R}}-m^2_{I'}}\right.\crn
&&\left.+\fr{s^2_{\theta_R}m^2_{R'_1}\ln \fr{m^2_{\nu_{\al R}}}{m^2_{R'_1}}}{m^2_{\nu_{\al R}}-m^2_{R'_1}}-\fr{s^2_{\theta_I}m^2_{I'_1}\ln \fr{m^2_{\nu_{\al R}}}{m^2_{I'_1}}}{m^2_{\nu_{\al R}}-m^2_{I'_1}}\right) \label{ketqua}.\eea     
It is noteworthy that the divergent parts arising from individual one-loop contributions by $R',I',R'_1,I'_1$, having a common form $C_{\mathrm{UV}}=\fr{1}{\ep}-\ga+\ln 4 \pi +1$ in dimensional regularization $\ep=2-d/2\to 0$, are exactly cancelled due to $(c^2_{\theta_R}-c^2_{\theta_I}+s^2_{\theta_R}-s^2_{\theta_I})C_{\mathrm{UV}}=0$.  Additionally, as suppressed by the loop factor $(1/32\pi^2)$, the $R',I'$ mass splitting $(m^2_{R'}-m^2_{I'})/m^2_{R',I'}\sim \la^2 v^2/w^2_{1,2}$, as well as the mixing angles $\theta^2_{R,I}\sim \la^2 v^2/w^2_{1,2}$, the resultant neutrino mass in (\ref{ketqua}) manifestly achieved, proportional to $m_{\nu}\sim \la^2 h^2v^2/32\pi^2 w_{1,2}$ is small, as expected.

Since $\eta$ is a mediator field, possibly coming from a more fundamental theory, I particularly assume it to be much heavier than other dark fields, i.e. $\mu_1\gg \mu_2, w_{1,2}$, thus $M_1\simeq \mu_1$ and one can take the soft coupling $\mu\lesssim \mu_1$. In this case, the diagram in Fig. \ref{fig1} approximates that in the basic scotogenic setup with the vertex $\fr 1 2 \bar{\la}(H\chi)^2+H.c.$ induced by $\eta$ to be $\bar{\la}=\la^2\mu w_2 w^2_1/\sqrt{2}\mu^4_1\sim (w_{1,2}/\mu_1)^3\ll 1$ explaining why $\bar{\la}$ is necessarily small \cite{scoto}. Indeed, it is clear that $\theta_R\simeq -\theta_I\simeq \la v w_1/2\mu_1^2$, commonly called $\theta=|\theta_{R,I}|$. The contribution of $R'_1, I'_1$ (i.e. $\eta$) in the last two terms in (\ref{ketqua}) is proportional to $\theta^2\ln m^2_{R'_1}/m^2_{I'_1}\simeq \bar{\la} v^2/\mu^2_1$, which is more suppressed than that by $\chi^0$ in the first two terms in (\ref{ketqua}) due to the $R',I'$ mass splitting, proportional to $(m^2_{R'}-m^2_{I'})/m^2_{R',I'}\simeq \bar{\la} v^2 /m^2_{R',I'}$, where notice that $m_{\nu_{\al R}}\sim m_{R',I'}\sim (\mu_2,w_{1,2}) \ll \mu_1$. That said, the neutrino mass is dominantly contributed by $R',I'$ (i.e. $\chi^0$), approximated as  
\be (m_\nu)_{ab}\simeq \fr{\bar{\la}v^2}{32\pi^2}\fr{h_{a\al}h_{b \al}  m_{\nu_{\al R}}}{m^2_{\chi^0}-m^2_{\nu_{\al R}}}\left(1-\fr{m^2_{\nu_{\al R}}}{m^2_{\chi^0}-m^2_{\nu_{\al R}}}\ln \fr{m^2_{\chi^0}}{m^2_{\nu_{\al R}}}\right),\ee because $m^2_{\chi^0}\equiv (m^2_{R'}+m^2_{I'})/2$ is much bigger than the $R',I'$ mass splitting. This matches the well-established result, but the smallness of the coupling $\bar{\la}$ or exactly of observed neutrino mass $m_{\nu}\sim 0.1$ eV given that $m_{\nu_{\al R}}\sim m_{\chi^0}\sim 1$ TeV and $h\sim 0.1$ is manifestly solved, since $\bar{\la}=(\la^2/\sqrt{2})(\mu/\mu_1) (w_2 w^2_1/\mu^3_1)\sim 10^{-6}$ for $w_{1,2}/\mu_1=10^{-2}$ and $\la\sim 1\sim \mu/\mu_1$, as desirable.       

{\it Dark matter}.---Differing from the scotogenic (odd) fields $\nu_{\al R},\chi^0,\eta$, the third right-handed neutrino ($\nu_{3R}$) is accidentally stabilized by the current model by itself despite the fact that it is even.\footnote{Such a minimal fermion dark matter was studied in \cite{Okada} in a $X$-charge setup, while a minimal scalar dark matter alternative to the minimal fermion dark matter was also given in \cite{Singirala} in a $B-L$ model.} Further, this stability is maintained even if $\nu_{3R}$ is heavier than the scotogenic fields and others. This results from the $B-L$ gauge symmetry solution for which $\nu_{3R}$ has a charge $B-L=5$ and is thus coupled only in pair in renormalizable interactions, say $\nu_{3R}\nu_{3R}\phi_2$ and $\bar{\nu}_{3R}\nu_{3R} Z'_{B-L}$. Given an effective interaction that leads to $\nu_{3R}$ decay, the one with minimal dimension is $\fr{1}{\La^3}\bar{l}_L\tilde{H}\nu_{3R}(\phi^2_1\phi_2)^*$, $\fr{1}{\La^3}\nu_{\al R} \nu_{3R} \eta (\phi^2_1\phi_2)^*$, or $\fr{1}{\La^5}\bar{l}_L\chi\nu_{3R}[\eta^*(\phi_1\phi_2)^2/\eta(\phi_1^3\phi_2)^*]$, where $\La\sim 10^{16}$~GeV would be a scale of GUT, broken for determining the effective couplings, conserving the current gauge symmetry. After $B-L$ breaking, $\nu_{3R}$ gets mass and possibly decays to normal fields $l_L H$, dark fields $\nu^c_{\al R} \eta^*$, or mixed product $l_L \chi^* \eta/\eta^*$, with the rate suppressed by $\Ga_{\nu_{3R}}\sim (w_{1,2}/\La)^{6} m_{\nu_{3R}}\to 0$ for the first two and $(w_{1,2}/\La)^{10} m_{\nu_{3R}}\to 0$ for the last one, since $w_{1,2}/\La\sim 10^{-13}$. It translates to a lifetime $\tau_{\nu_{3R}}\sim 10^{43} (\mathrm{TeV}/m_{\nu_{3R}})(\La/10^{13}w_{1,2})^6\ \mathrm{yr}\sim 10^{43}\ \mathrm{yr}$ and $10^{95} (\mathrm{TeV}/m_{\nu_{3R}})(\La/10^{13}w_{1,2})^{10}\ \mathrm{yr}\sim 10^{95}\ \mathrm{yr}$, given that $m_{\nu_{3R}}\sim$ TeV, respectively. It indicates that the field $\nu_{3R}$ is absolutely stable. 

\begin{figure}[h]
\bc
\includegraphics[scale=1]{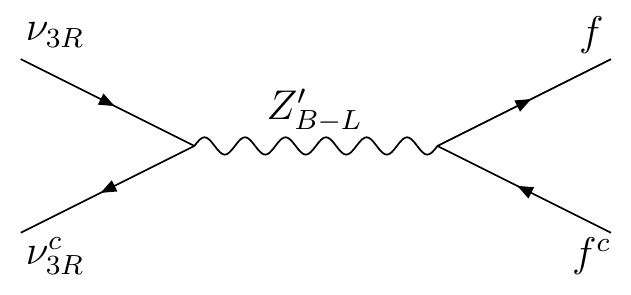}
\caption[]{\label{fig2} Annihilation of accidental $\nu_{3R}$ dark matter to normal matter and possible odd field.}
\ec
\end{figure}       

The field $\nu_{3R}$ communicates with normal matter through the $Z'_{B-L}$ and $\phi_2$ portals only, unlike the scotogenic fields that interact directly with usual leptons and Higgs field additionally. Obviously the $\phi_2$ portal couples to normal matter only through a mixing with the usual Higgs field, giving a small contribution to dark matter observables. The gauge portal dominantly contributes to dark matter annihilation to normal matter via $s$-channel diagrams as in Fig. \ref{fig2} where $f$ is every fermion, possibly including odd field. The annihilation cross-section is proportional to $\langle \sigma v_{\mathrm{rel}} \rangle_{\nu_{3R}} \sim g^4_{B-L}m^2_{\nu_{3R}}/(4m^2_{\nu_{3R}}-m^2_{Z'_{B-L}})^2$. Hence, the $Z'$ mass resonance $m_{\nu_{3R}}\simeq \fr 1 2 m_{Z'_{B-L}}$ will set a correct relic density for dark matter, i.e. $\Om_{\nu_{3R}} h^2\simeq 0.1\ \mathrm{pb}/\langle \sigma v_{\mathrm{rel}}\rangle_{\nu_{3R}}\leq 0.12$ \cite{pdg}. Let us remind the reader that throughout this work ``$h$'' that is {\it always} coupled to the density ``$\Om$'' denotes the reduced Hubble parameter, without confusion with the Yukawa coupling $h_{a\al}$ or even $h$'s when the indices ($a\al$) are suppressed. Note that $\nu_{3R}$ is a Majorana field, scattering with nucleon in direct detection only via spin-dependent effective interaction through exchange by $Z'_{B-L}$. However, this kind of interaction of $Z'_{B-L}$ with quarks confined in nucleon vanishes, hence it presents a negative search result, as currently measured \cite{dddsd}.

\begin{figure}[h]
\bc
\includegraphics[scale=1]{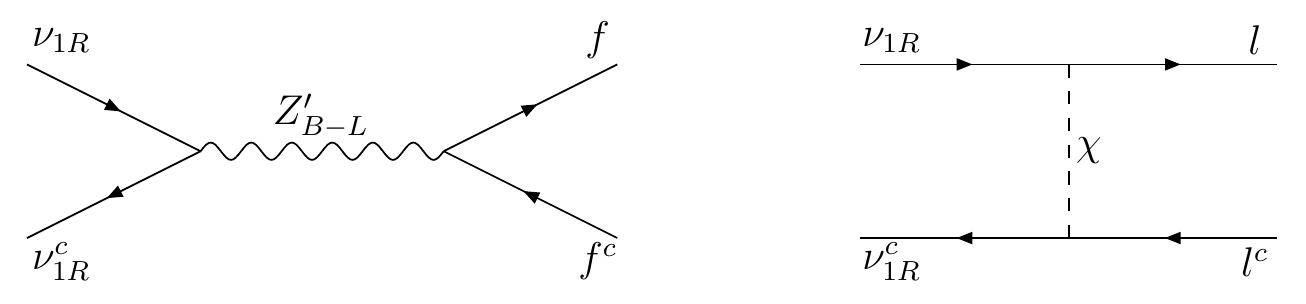}
\caption[]{\label{fig3} Annihilation of scotogenic $\nu_{1R}$ dark matter to normal matter and possible $\nu_{3R}$ field.}
\ec
\end{figure}       

Last, but not least, the lightest of odd fields $\nu_{\al R}, \chi^0, \eta$ is stabilized by the matter parity, potentially contributing to dark matter too. As a result, this model presents a promising scenario for two-component dark matter \cite{mcdm}. In what follows, I choose the lightest odd field to be $\nu_{1R}$ and interpret the world of dark matter with two right-handed neutrino components, $\nu_{1R}$ and $\nu_{3R}$ above. First note that due to Majorana nature, $\nu_{1,3R}$ annihilation is helicity-suppressed. Hence, it is important to include $p$-wave contributions to their annihilation cross-section too. In the early universe, the field $\nu_{3R}$ annihilates to usual fermions $(l,q)$ as well as the field $\nu_{1R}$ if $m_{\nu_{3R}}>m_{\nu_{1R}}$ via the diagrams identical to Fig.~\ref{fig2} for $f=l,q,\nu_{1R}$, revealing an annihilation cross-section,
\bea \langle \sigma v_{\mathrm{rel}} \rangle_{\nu_{3R}} &=& \fr{325 \langle v^2_{\mathrm{rel}} \rangle g^4_{B-L} m^2_{\nu_{3R}}}{6\pi (4m^2_{\nu_{3R}}-m^2_{Z'_{B-L}})^2}+\Theta(m_{\nu_{3R}}-m_{\nu_{1R}}) \crn
&&\times \fr{200 g^4_{B-L}}{\pi} \sqrt{1-\fr{m^2_{\nu_{1R}}}{m^2_{\nu_{3R}}}} \left[\fr{m^2_{\nu_{1R}}}{2 m^4_{Z'_{B-L}}}+\fr{\langle v^2_{\mathrm{rel}}\rangle (m^2_{\nu_{3R}} - m^2_{\nu_{1R}})}{3 (4m^2_{\nu_{3R}}-m^2_{Z'_{B-L}})^2}\right].\eea On the other hand, the field $\nu_{1R}$ annihilates to usual fermions as well as the field $\nu_{3R}$ if $m_{\nu_{1R}}>m_{\nu_{3R}}$ via diagrams as in Fig. \ref{fig3} for $f=l,q,\nu_{3R}$. It is straightforward to compute $\nu_{1R}$ annihilation cross-section summarizing all $t,s$-channel diagrams, such as 
\bea \langle \sigma v_{\mathrm{rel}} \rangle_{\nu_{1R}} &=& \fr{|h|^4}{32\pi}\fr{m^2_{\nu_{1R}}}{(m^2_{\nu_{1R}}+m^2_\chi)^2} +\fr{5\langle v^2_{\mathrm{rel}}\rangle |h|^2 g^2_{B-L}m^2_{\nu_{1R}}}{6\pi(m^2_{\nu_{1R}}+m^2_\chi)(4m^2_{\nu_{1R}}-m^2_{Z'_{B-L}})}+ \fr{104 \langle v^2_{\mathrm{rel}}\rangle g^4_{B-L} m^2_{\nu_{1R}}}{3\pi (4 m^2_{\nu_{1R}}-m^2_{Z'_{B-L}})^2}\crn
&&+\Theta(m_{\nu_{1R}}-m_{\nu_{3R}}) \fr{200 g^4_{B-L}}{\pi} \sqrt{1-\fr{m^2_{\nu_{3R}}}{m^2_{\nu_{1R}}}}\left[\fr{m^2_{\nu_{3R}}}{2m^4_{Z'_{B-L}}}+\fr{\langle v^2_{\mathrm{rel}}\rangle (m^2_{\nu_{1R}}- m^2_{\nu_{3R}})}{3(4m^2_{\nu_{1R}}-m^2_{Z'_{B-L}})^2}\right]. \eea It is clear that for $\nu_{1R}$ the $t$-channel diagram gives a contribution in addition to the $s$-channel diagram, while it does not exist for the case of $\nu_{3R}$. The $t$-channel contribution arises from six physical odd fields $\chi\in(\chi^\pm,R',R'_1,I',I'_1)$, which must be heavier than $\nu_{1R}$. Since the $\nu_{1R}$ relic density is necessarily governed by $Z'_{B-L}$ mass resonance, I suppose the $t$-channel contribution effectively mediated by a characteristic particle $\chi$ that has a mass $m_\chi \sim m_{Z'_{B-L}}>m_{\nu_{1R}}$ with effective coupling signified as $h\sim g_{B-L}$. Notice that the conversion $\nu_{1R}\leftrightarrow \nu_{3R}$ in either annihilation may get a contribution of $\phi_{1,2}$ besides $Z'_{B-L}$, but this effect is negligible if the $\phi_{1,2}$ fields small mix, as supposed. 

\begin{figure}[h]
\bc
\includegraphics[scale=0.5]{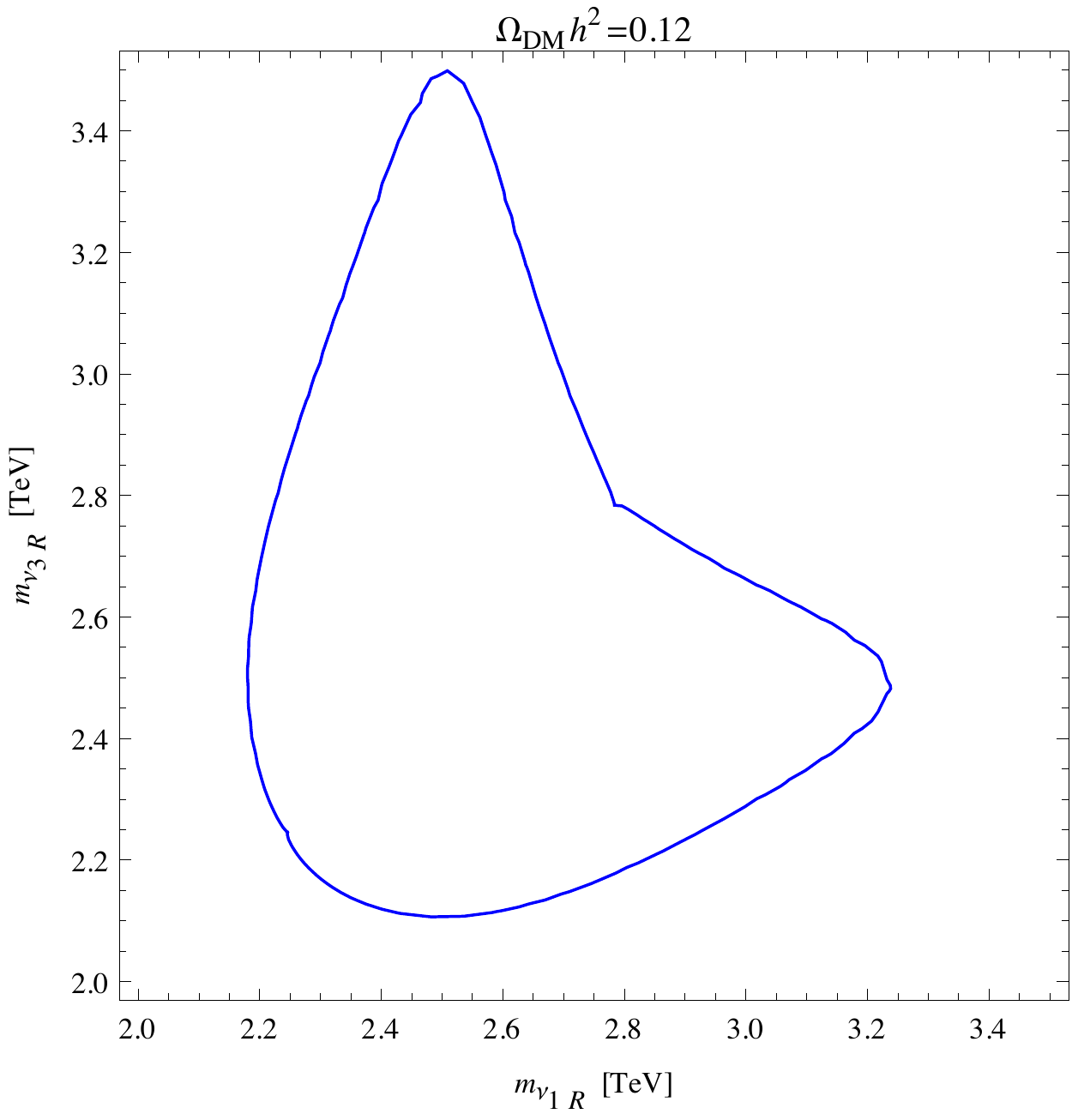}
\includegraphics[scale=0.5]{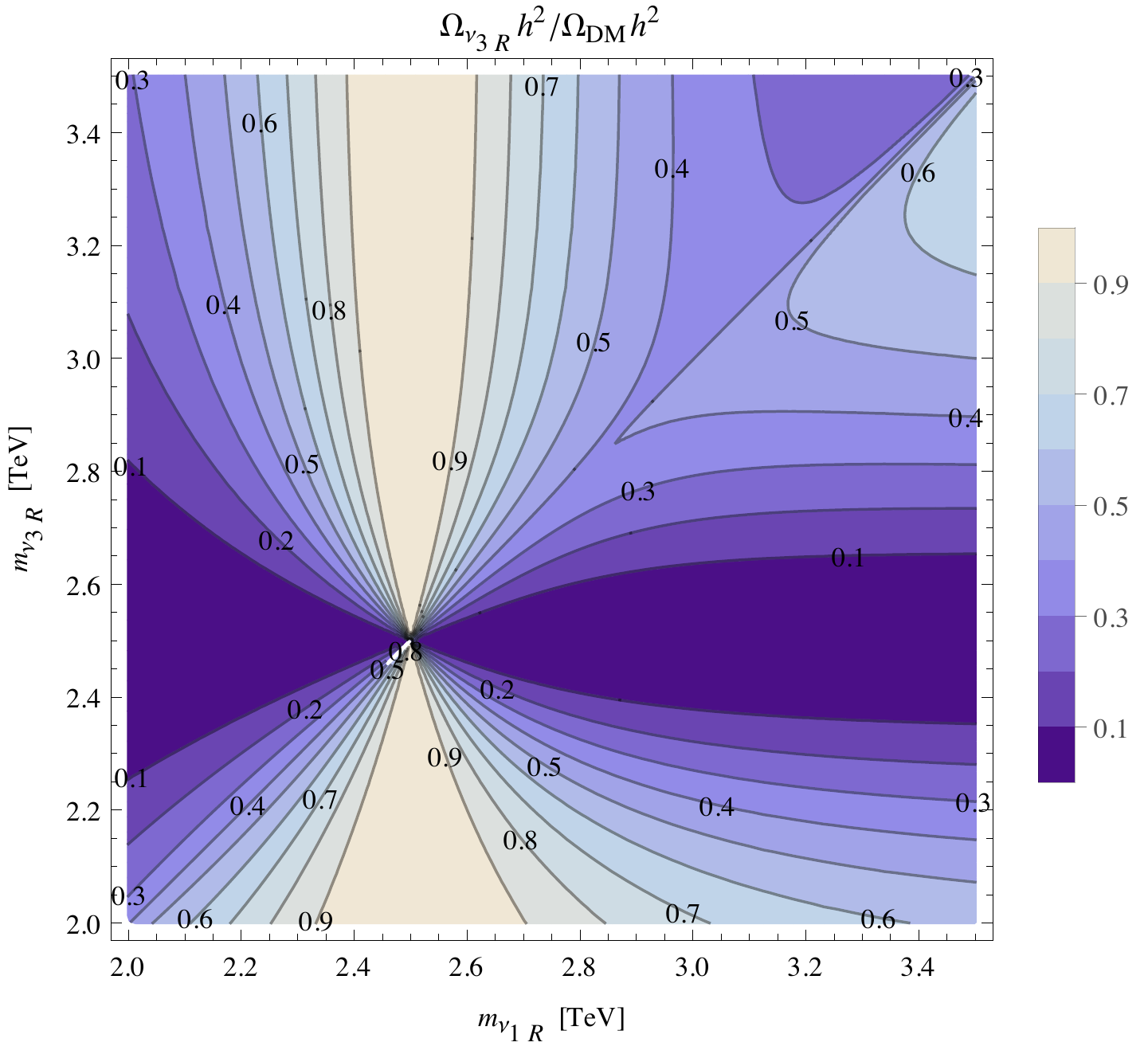}
\caption[]{\label{fig4} Total relic density of $\nu_{1,3R}$ (left panel) as well as contribution of $\nu_{3R}$ to the total relic density (right panel) contoured as function of their masses.}
\ec
\end{figure}            

The densities of $\nu_{1,3R}$ are obtained by solving the coupled Boltzmann equations which contain their annihilation to standard model fields and conversion between themselves \cite{dong}. When generation of lighter dark matter $\nu_{1R}$($\nu_{3R}$) from heavier dark matter $\nu_{3R}$($\nu_{1R}$) is less significant compared to its annihilation to usual fermions, the approximate analytic solution is given by $\Om_{\nu_{1R}}h^2\simeq 0.1\ \mathrm{pb}/\langle \sigma v_{\mathrm{rel}}\rangle_{\nu_{1R}}$ and $\Om_{\nu_{3R}}h^2\simeq 0.1\ \mathrm{pb}/\langle \sigma v_{\mathrm{rel}}\rangle_{\nu_{3R}}$. I make a contour of the total relic density $\Om_{\mathrm{DM}} h^2 = \Om_{\nu_{1R}} h^2+\Om_{\nu_{3R}} h^2=0.12$---where the last value is experimentally measured---as function of $(m_{\nu_{1R}},m_{\nu_{3R}})$, according to $m_\chi=m_{Z'_{B-L}}=5$ TeV for fixed $h=g_{B-L}=0.28$ (see below) and $\langle v^2_{\mathrm{rel}}\rangle =6/x_F$ with $x_F=25$ for each dark matter component, as in Fig. \ref{fig4} left panel. To see the contribution of each component to the total density, the ratio $\Om_{\nu_{3R}}h^2/\Om_{\mathrm{DM}}h^2$ is also contoured in Fig. \ref{fig4} right panel (contribution of $\nu_{1R}$ is followed by $1-\Om_{\nu_{3R}}h^2/\Om_{\mathrm{DM}}h^2$, which is not plotted). For each value of $m_{Z'_{B-L}}$, the relevant mass resonances $m_{\nu_{1R}}=\fr 1 2 m_{Z'_{B-L}}$ (as vertical line) and $m_{\nu_{3R}}=\fr 1 2 m_{Z'_{B-L}}$ (as horizontal line) are crucial to set the correct relic density $\Om_{\mathrm{DM}} h^2=0.12$ as the density curve is based/distributed around these resonant lines. Additionally, if the mass resonance occurs at $\nu_{1R}$ then its partner $\nu_{3R}$ mainly contributes to the density, and vice versa.  Lastly, as $\nu_{3R}$ in previous scenario, both $\nu_{1,3R}$ in two-component dark matter scheme possess a negligible scattering cross-section with nuclei in direct detection, appropriate to observation.  

{\it Concerning collider limits}.---$Z'_{B-L}$ couples to both leptons and quarks, presenting promising signals at colliders. The LEPII experiment \cite{lepii} searched for such a new gauge boson through process $e^+e^-\to f\bar{f}$ for $f=\mu,\tau$, described by the effective Lagrangian $\mathcal{L}_{\mathrm{eff}}\supset (g_{B-L}/m_{Z'_{B-L}})^2(\bar{e}\ga^\mu e)(\bar{f} \ga_\mu f)$, making a bound $m_{Z'_{B-L}}/g_{B-L}>6$ TeV. Since $m_{Z'_{B-L}}=g_{B-L}\sqrt{64w^2_1+100w^2_2}$, it correspondingly limits $\sqrt{64 w^2_1+100 w^2_2}>6$ TeV; particularly, $w_{1}\sim w_{2}\gtrsim 0.5$ TeV if the two scales are equivalent. Alternatively, the LHC experiment \cite{lhc1,lhc2} looked for dilepton signals via process $pp\to Z'_{B-L}\to f\bar{f}$ for $f=e,\mu$, yielding a $Z'_{B-L}$ mass bound roundly $m_{Z'_{B-L}}\gtrsim 5$~TeV for $Z'_{B-L}$ coupling relative to that of $Z$, such as $g_{B-L}=\sqrt{5/8} s_W g_Z\simeq 0.28$. This converts to $\sqrt{64 w^2_1+100 w^2_2}= m_{Z'_{B-L}}/g_{B-L}\gtrsim 17.85$ TeV, thus $w_1\sim w_2\gtrsim 1.39$ TeV, which is radically bigger than the LEPII. The last bound is appropriate to those imposed for neutrino mass and dark matter, as desirable.     

{\it Concluding remarks}.---The dark side of the $B-L$ gauge symmetry is perhaps associated with three right-handed neutrinos that possess $B-L=-4,-4,+5$, respectively. This theory implies a unique matter parity as residual gauge symmetry, stabilizing scotogenic fields in a way different from the hypothesis of superparticles. Besides explaining the scotogenic neutrino mass generation and dark matter candidate, the model reveals a second component for dark matter, $\nu_{3R}$ with $B-L=5$.


\begin{thebibliography}{99}
 
\bibitem{blnumber} J. C. Montero and V. Pleitez, Phys. Lett. B {\bf 675}, 64 (2009).

\bibitem{Minkowski} P.~Minkowski, Phys. Lett. B {\bf 67}, 421 (1977).

\bibitem{Yanagida} T.~Yanagida, Conf. Proc. C {\bf 7902131}, 95 (1979).

\bibitem{Gell-Mann} M.~Gell-Mann, P.~Ramond, and R.~Slansky, Conf. Proc. C {\bf 790927}, 315 (1979).

\bibitem{Mohapatra} R.~N.~Mohapatra and G.~Senjanovic, Phys. Rev. Lett. {\bf 44}, 912 (1980).

\bibitem{addbl1} J. C. Montero and B. L. Sanchez-Vega, Phys. Rev. D {\bf 84}, 053006 (2011).

\bibitem{addbl2} B. L. Sanchez-Vega, J. C. Montero, and E. R. Schmitz, Phys. Rev. D {\bf 90}, 055022 (2014).

\bibitem{Okada} N.~Okada, S.~Okada, and D.~Raut, Phys. Rev. D {\bf 100}, 035022 (2019).

\bibitem{Singirala} S.~Singirala, R.~Mohanta, and S.~Patra, Eur. Phys. J. Plus {\bf 133}, 477 (2018).

\bibitem{addbl5} E. Ma and R. Srivastava, Phys. Lett. B {\bf 741}, 217 (2015).

\bibitem{addbl6} E. Ma, N. Pollard, R. Srivastava, and M. Zakeri, Phys. Lett. B {\bf 750}, 135 (2015). 

\bibitem{addblno} T. Nomura and H. Okada, Eur. Phys. J. C {\bf 78}, 189 (2018).

\bibitem{addbl7} H. Okada, Y. Orikasa, and Y. Shoji, JCAP {\bf 07}, 006 (2021).

\bibitem{Ma} E.~Ma, N.~Pollard, O.~Popov, and M.~Zakeri, Mod. Phys. Lett. A {\bf 31}, 1650163 (2016).

\bibitem{addbl9} S. Mishra, N. Narendra, P. K. Panda, and N. Sahoo, Nucl. Phys. B {\bf 981}, 115855 (2022).

\bibitem{scoto} E. Ma, Phys. Rev. D {\bf 73}, 077301 (2006).

\bibitem{matterparity} P. V. Dong, C. H. Nam, and D. V. Loi, Phys. Rev. D {\bf 103}, 095016 (2021).

\bibitem{copositivity} K. Kannike, Eur. Phys. J. C {\bf 76}, 324 (2016). [Erratum: Eur. Phys. J. C {\bf 78}, 355 (2018)].

\bibitem{pdg} R. L. Workman {\it et al.} (Particle Data Group), Prog. Theor. Exp. Phys. {\bf 2022}, 083C01 (2022).

\bibitem{dddsd} J. Aalbers {\it et al.} (LUX-ZEPLIN Collaboration), Phys. Rev. Lett. {\bf 131}, 041002 (2023), arXiv:2207.03764 [hep-ex].

\bibitem{mcdm} See, e.g., C. Boehm, P. Fayet, and J. Silk, Phys. Rev. D {\bf 69} 101302 (2004). 

\bibitem{dong} C. H. Nam, D. V. Loi, L. X. Thuy, and P. V. Dong, JHEP {\bf 12}, 029 (2020).

\bibitem{lepii} J. Alcaraz {\it et al.} (ALEPH, DELPHI, L3, OPAL Collaborations, and LEP Electroweak Working Group), arXiv:hep-ex/0612034.

\bibitem{lhc1} G. Aad {\it et al.} (ATLAS Collaboration), Phys. Lett. B {\bf 796}, 68 (2019).

\bibitem{lhc2} A. M. Sirunyan {\it et al.} (CMS Collaboration), JHEP {\bf 07}, 208 (2021).


\end{thebibliography}
\end{document}